\documentclass[journal,final]{IEEEtran}
\usepackage[T1]{fontenc}
\usepackage{bm}
\usepackage{amsmath}
\usepackage{amsthm}
\usepackage{amssymb}
\usepackage{graphicx}

\usepackage{stfloats}
\usepackage{subfigure} 
\usepackage{booktabs}
\usepackage{multicol}
\usepackage{multirow}
\usepackage{makecell}
\usepackage{color}

\usepackage[unicode=true,
bookmarks=true,bookmarksnumbered=true,bookmarksopen=true,bookmarksopenlevel=1,
breaklinks=false,pdfborder={0 0 0},pdfborderstyle={},backref=false,colorlinks=false]
{hyperref}
\hypersetup{pdftitle={Your Title},
	pdfauthor={Your Name},
	pdfpagelayout=OneColumn, pdfnewwindow=true, pdfstartview=XYZ, plainpages=false}

\makeatletter
\theoremstyle{plain}
\newtheorem{thm}{\protect\theoremname}
\theoremstyle{definition}

\theoremstyle{remark}

\theoremstyle{plain}

\theoremstyle{plain}

\usepackage[vlined,boxed,ruled,linesnumbered,resetcount]{algorithm2e}
\usepackage{algpseudocode}
\usepackage[noadjust,sort,compress]{cite}

\providecommand{\definitionname}{Definition}
\providecommand{\lemmaname}{Lemma}
\providecommand{\propositionname}{Proposition}
\providecommand{\remarkname}{Remark}
\providecommand{\theoremname}{Theorem}
\usepackage[justification=centering]{caption}

\usepackage{color, colortbl}
	
\definecolor{Gray}{gray}{0.9}
\usepackage[first=0,last=9]{lcg}

\makeatother
 
\begin{document}
\title{Lattice-Based Minimum-Distortion Data
	Hiding}

\author{Jieni Lin, Junren Qin, Shanxiang Lyu, Bingwen Feng, Jiabo Wang
	\thanks{This work was supported partly by 
		the Major Program of Guangdong Basic and Applied Research  (Grant No. 2019B030302008),
		the Key R\&D Program of Guangdong Province (Grant No. 2019B010136003), National Natural Science Foundation of China (Grant No. 61902149, 62032009, 61932010, and 61802145), the Natural Science Foundation of Guangdong Province (Grant No. 2020A1515010393 and 2019B010137005), Science and Technology Program of Guangzhou, China (Grant No. 202007040004 and 201804010428), and the Fundamental Research Funds for the Central Universities (Grant No. 21620438).}
	\thanks{
		Jieni Lin, Junren Qin, Shanxiang Lyu and Bingwen Feng are with the College of Cyber Security, Jinan
		University, Guangzhou 510632, China (e-mail: jennie000@stu2018.jnu.edu.cn; alex2abee@stu2020.jnu.edu.cn;  shanxianglyu@gmail.com; bingwfeng@gmail.com). 
	Shanxiang Lyu is also with the State
		Key Laboratory of Cryptology, P.O. Box 5159, Beijing, 100878, China.
Jiabo Wang is with the Beijing National Research Center for Information Science and Technology, Tsinghua University, Beijing 100084, China (e-mail: wangjiabo@mail.tsinghua.edu.cn).
		Jieni Lin and Junren Qin contributed equally to this work.  \textit{(Corresponding author: Shanxiang Lyu.)}
	}
}
\maketitle

\begin{abstract}
Lattices  have been conceived as a powerful tool for data hiding. While conventional studies and applications focus on achieving the optimal robustness versus distortion tradeoff, in some applications such as  data hiding in medical/physiological signals, the primary concern is to achieve a minimum amount of distortion to the cover signal. In this paper, we revisit the  celebrated quantization index modulation (QIM)  scheme and propose a minimum-distortion version of it, referred to as MD-QIM.
The crux of MD-QIM is to move the data point to only the boundary of the Voronoi region of the lattice point indexed by a message, which suffices for subsequent correct decoding. At any fixed code rate, the scheme achieves the minimum amount of distortion by sacrificing the robustness to the additive white Gaussian noise (AWGN) attacks.
 Simulation results confirm that our scheme significantly outperforms QIM
 in terms of mean square error (MSE), peak signal to noise ratio (PSNR) and percentage residual difference (PRD).
\end{abstract}

\section{Introduction}
\noindent
\IEEEPARstart{D}{ata} hiding is to embed one signal (referred to as a ``message'' or ``watermark''), to within  another signal (referred to as a ``cover'' or ``host signal''). The embedding should be unnoticeable,  \textit{i.e.}, causing no serious degradation to the cover.  
Since Cox \textit{et al.} \cite{cox99} pointed out the close relationship between
data hiding and communication over a channel with
side information, guided by well known information-theoretic results \cite{zamir2014,lingtit14}, the focus of the subject has shifted to designing practical schemes for  steganography, digital watermarking,  covered communications, and so on \cite{feng16a,feng16b,changshengtip19}.
 
In the data hiding community, great attention has been given to the class of quantization index modulation (QIM) algorithms \cite{ChenW01}, in which 
  the host signal is quantized to the nearest point of the codebook indexed by messages \cite{MoulinK05}. 
 A critical problem in QIM is to design good codebooks which are robust against the additive white Gaussian noise (AWGN) attacks. 
 In this regard, researchers have resorted to using  optimal low-dimensional lattice-based quantizers  where efficient encoding and decoding algorithms exist \cite{e803,conway2013sphere}. For instance, Zhang and Boston \cite{e803} employed $E_8$ lattices to arrive at an $E_8$-based QIM. To date many   
 QIM variants have been proposed based on various characteristics.
  For example, adopting a set of scalar and uniform quantizers results in the popular Dither Modulation (DM) scheme and its Distortion Compensated version (DC-DM) \cite{ChenW01}, changing the domain of real numbers to the angular domain and logarithmic domain results to angle QIM \cite{OuriqueLJP05} and logarithmic quantization \cite{KalantariA10}. 
     In addition, the method \cite{ibaida2011low} that combines QIM and least significant bit (LSB) embedding has enabled high capacity Electrocardiography (ECG) signal watermarking, which outperforms the chirp-signal based method \cite{kaur2010digital}.

 {
As a blind approach, QIM can achieve provably better rate-distortion–robustness tradeoffs \cite{ChenW01}.} Nevertheless, there are also some applications where the robustness to the AWGN attack is not our primary concern. For instance, the {ECG} steganography in Point-of-Care  systems \cite{rmit1} is to hide data collected from body sensors inside an ECG signal such that a
person in the middle cannot even detect its existence. Generally the attack is passive.  {Hence the data hiding technique should  guarantee a minimum acceptable distortion in the
	ECG signal, such that the watermarked ECG would not affect the accuracy of further diagnoses.} Against this background, for any specified information rate,  {how to tweak the conventional QIM scheme towards a version of minimum distortion becomes an interesting open problem.}

Addressing the minimum-distortion issue, the contributions of this paper are summarized as follows:

\begin{itemize}
	\item  Firstly, we reformulate the information embedding and extraction processes of QIM from the perspective of coset coding.  {Noticing that an embedded data point is not necessarily coerced} to a lattice point for the sake a correct decoding, but rather any point inside the Voronoi region suffices, we propose an improved QIM scheme (referred to as MD-QIM) in which the cover signal is either staying still, or only moved to the border of a packing region. This {novel} design is feasible for any lattice basis and for any specified information rate.
	\item  Secondly, we rigorously analyze the mean square error (MSE) difference between MD-QIM and the original QIM. Together with the known MSE caused by QIM, the MSE of MD-QIM becomes available. Simulation results show that our lower bound on MSE is quite tight. 
	\item  Thirdly, we justify the performance advantages of MD-QIM by numerical simulations based on ECG signals.  Considering the metrics of mean square error (MSE), peak signal to noise ratio (PSNR) and percentage residual difference (PRD), our simulation results show that MD-QIM outperforms the conventional QIM significantly.
\end{itemize}
 

\section{Preliminaries}


\subsection{Lattices}
An $N$-dimensional lattice in $\mathbb{R}^N$ is a discrete additive subgroup $\Lambda=\{\mathbf{G}\mathbf{z}|\mathbf{z} \in \mathbb{Z}^N\}$, where the full-ranked matrix $\mathbf{G} \in \mathbb{R}^{N\times N}$ is called the generator matrix of $\Lambda$. 
The nearest neighbor quantizer $Q_\Lambda(\cdot)$ \textit{w.r.t.} $\mathbf{x}\in \mathbb{R}^N$ is defined as
\begin{align}
  Q_\Lambda(\mathbf{x})=\mathop{\arg\min}_{\mathbf{\lambda} \in \Lambda} \|\mathbf{x}-\mathbf{\lambda}\|.
\end{align}
The Voronoi cell $\mathcal{V}_\lambda$ of a lattice point $\lambda \in \Lambda$ is the set of points in $\mathbb{R}^{N\times N}$ which are quantized to $\lambda$:
\begin{align}
\mathcal{V}_\lambda=\{\mathbf{x}:Q_\Lambda(\mathbf{x})=\lambda\},
\end{align}
where the Voronoi region for the zero vector is referred to as the fundamental Voronoi region
\begin{align}
\mathcal{V}_\Lambda=\{\mathbf{x}:Q_\Lambda(\mathbf{x})=\mathbf{0}\}.
\end{align}
The volume of the fundamental region is defined by
\begin{align}
    \rm{Vol}(\mathcal{V}_\Lambda)=\int_{\mathcal{V}_{\Lambda}} d\mathbf{x}=|\rm{det} \mathbf{G}|.
\end{align}
The normalized second moment  of a lattice $\Lambda$ is defined by
\begin{equation}
	G(\Lambda)=\frac{1}{\rm{Vol}(\mathcal{V}_\Lambda)^{\frac{2}{N}}} \times \frac{\int_{\mathbf{x}\in \mathcal{V}_\Lambda } ||\mathbf{x}||^2 \mathrm{d}\mathbf{x}}{N \rm{Vol}(\mathcal{V}_\Lambda)}.
\end{equation}
The packing radius of $\Lambda$ is given by
\begin{align}
r_{\rm{pack}(\Lambda)} = \frac{1}{2} d_{\rm{min}}(\Lambda) {,}
\end{align}
where $d_{\rm{min}}(\Lambda) = \rm{min}_{\lambda \in \Lambda \backslash \{\mathbf{0}\}}\|\lambda\|$ denotes minimum Euclidean distance between any two lattice points.

We say that two lattices $\Lambda_f$ and $\Lambda_c$ are nested if $\Lambda_c\subset \Lambda_f$. The bigger lattice $\Lambda_f$ is called the \textit{fine/coding} lattice, and $\Lambda_c$ is called the \textit{coarse/shaping} lattice.
A typical and efficient method to build a set of lattice points for data hiding, called coset coding, is to use nested lattices based shaping \cite{MoulinK05,zamir2014}.

\subsection{Encoding of QIM}
\noindent \textbf{Step i)} Consider a host signal vector $\mathbf{s}$ and a set of messages $\mathcal{M}$ to be  {embedded}.
$\mathcal{I}$ is the set of  {indices} of messages.
The message $\mathbf{m}_i \in \mathcal{M}$ can be recognized through the index $i \in \mathcal{I}$.

\noindent \textbf{Step ii)} Consider two nested lattice $\Lambda_f$ and $\Lambda_c \subset \Lambda_f$ in $\mathbb{R}^N$.  The generator matrices of $\Lambda_c$ and $\Lambda_f$  {which are denoted by 
 $\mathbf{G}_c$ and $\mathbf{G}_f$, respectively, have the following relation}
 \begin{align}
     \mathbf{G}_c = \mathbf{G}_f \mathbf{J} {,}
 \end{align}
 where $\mathbf{J}$ is a subsampling matrix.
 The fine lattice $\Lambda_f$ can be decomposed as the union of $|\rm{det} \mathbf{J}|$ cosets of the coarse lattice $\Lambda_c$:
  {\begin{align}
 	\Lambda_f = \bigcup\limits_{i=0}^{|\rm{det} \mathbf{J}|-1} \Lambda_i
 	=\bigcup\limits_{\mathbf{d}_i \in \Lambda_f  {\slash} \Lambda_c} (\mathbf{d}_i + \Lambda_c),
 	\end{align}}
where each coset $\Lambda_i = \Lambda_c  + \mathbf{d}_i$ is a translated coarse lattice and $\mathbf{d}_i$ is called the coset representative of $\Lambda_i$. 

\noindent \textbf{Step iii)} To hide $\mathbf{m}_i$ in $\mathbf{s}$, the  QIM encoder quantizes $\mathbf{s}$ to the nearest point in $\Lambda_i$ by
\begin{equation} \label{QIM_embed}
\mathbf{s}_w \triangleq	Q_{\Lambda_i}(\mathbf{s}) = Q_{\Lambda_c}(\mathbf{s}-\mathbf{d}_i)+\mathbf{d}_i.
\end{equation}
The index $i$ in $\Lambda_i$ 
indicates that this coset is used for transmitting 
the massage $\mathbf{m}_i$. Moreover,
the payload of this embedding is $|\rm{det} \mathbf{J}|$, and the code rate is 
\begin{equation}
	R=\frac{1}{N}\log {|\rm{det} \mathbf{J}|}.
\end{equation}

\subsection{Decoding of QIM}
Assume that  the received signal $\mathbf{y}$ has been contaminated by a noise term $\mathbf{n}$: $\mathbf{y}= \mathbf{s}_w + \mathbf{n}$.
To extract the embedded message $\mathbf{m}_i$ from $\mathbf{y}$, the QIM decoder searches for  the closest coset $\Lambda_i$ to estimate the message index:  
\begin{equation}\label{qim_decode}
	  \hat{i}=\mathop{\arg\min}_{i \in \{0, 1, \dots, |\rm{det} \mathbf{J}|-1\}}
	{\rm{dist}}(\mathbf{y},\Lambda_i),
\end{equation}
where dist$(\mathbf{y},\Lambda)\triangleq \min_{\mathbf{\lambda} \in \Lambda}\|\mathbf{y}-\mathbf{\lambda}\|$.
If the noise is small enough such that $Q_{\Lambda_f}(\mathbf{n})=\mathbf{0}$, then the
 extracted message $\mathbf{m}_{\hat{i}}$ is correct.

\section{The Proposed Method}
 {In the passive steganography scenario where there is no AWGN attack, moving the host signal to any point inside the Voronoi cell suffices for subsequent decoding. }
Based on this idea, we propose an MD-QIM scheme which     achieves a minimum amount of distortion within the QIM framework. 
 {MD-QIM does not quantize the cover to a lattice point, but rather to a point in the boarder of the Voronoi region.
	In this section, we describe the   construction of MD-QIM, followed by its theoretical analysis.}

\subsection{MD-QIM}



In terms of encoding, MD-QIM also employs Steps i) and ii) in Section II-B) to prepare the cosets. The Step iii) in MD-QIM is more delicate. 
Specifically, our Step iii) consists of the following procedures:

\noindent \textbf{Step iii-1)} Employ Eq. (\ref{QIM_embed}) to calculate
\begin{equation}
	\mathbf{x}_i \triangleq Q_{\Lambda_i}(\mathbf{s})=Q_{\Lambda_c}(\mathbf{s}-\mathbf{d}_i)+\mathbf{d}_i.
\end{equation} 

\noindent   {\textbf{{Step iii-2)}} Calculate the  {difference vector} $\mathbf{p}_i$ between the lattice point and the cover:
	\begin{align}
	\mathbf{p}_i = \mathbf{x}_i - \mathbf{s}.
	\end{align}
	If  {$\| \mathbf{p}_i \| < r_{\rm{pack}(\Lambda_f)}$}, then the final embedded point is set as \begin{equation} \label{staystill}
	\mathbf{s}_w = \mathbf{s}. \, (\mathrm{Type\,I})
	\end{equation}
	Otherwise, continue with the steps below.}

\noindent \textbf{{Step iii-3)}} The final embedded point $\mathbf{s}_w$ is set as
\begin{align} \label{cost_mdqim_2}
    \mathbf{s}_w = \mathbf{x}_i - 
    \frac{\mathbf{p}_i}{\| \mathbf{p}_i \|}\times(r_{\rm{pack}(\Lambda_f)} - \epsilon), \,(\mathrm{Type\,II})
\end{align}
where $\epsilon\rightarrow 0$ is a small positive number to move $\mathbf{s}_w$ away from the decision boarders.

In terms of decoding, MD-QIM also uses Eq. (\ref{qim_decode}) to estimate the embedded message $\mathbf{m}_i$. Since 
\begin{equation}
	Q_{\Lambda_f}\left(\frac{\mathbf{p}_i}{\| \mathbf{p}_i \|}\times(r_{\rm{pack}(\Lambda_f)} - \epsilon)\right)=\mathbf{0},
\end{equation}
correct decoding is enabled.

\subsection{Workout Example}
\begin{figure}[t!]
	\centering{\includegraphics[width=.38\textwidth]{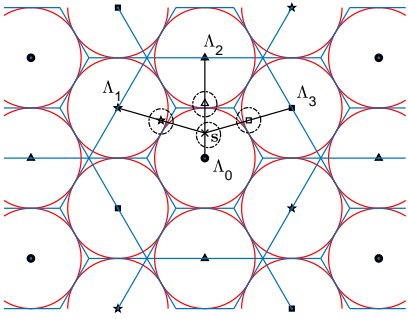}}
	\caption{The rationale of encoding in MD-QIM, where the encoded points are marked with dash circles. 
	}\label{figCase}
\end{figure}
For very large $N$, lattices which are simultaneously good for AWGN and quantization \cite{lyutit19} can be employed to design optimal $\Lambda_f$ and $\Lambda_c$ for both QIM and MD-QIM. For practical small $N$, nested lattices can be devised from celebrated low-dimensional lattices such as $A_2$, $D_4$, $E_8$ \textit{etc}.
Hereby we choose the $A_2$ lattice as $\Lambda_f$, and the self-nested $2 \Lambda_f$  as $\Lambda_c$  to illustrate the construction of MD-QIM.


The $A_2$ lattice has a generator matrix 
\begin{equation}
\mathbf{G}=\left(\begin{smallmatrix}
0 & \frac{\sqrt{3}}{2} \\ 1 & \frac{1}{2} \end{smallmatrix}\right).
\end{equation}
The $4$ cosets which correspond to $\mathcal{I}=\{0, 1, 2, 3\}$ are respectively given by
\begin{align*}
\Lambda_0 &=\mathbf{G}([0,0]^\top+\mathbb{Z}^2), \\
\Lambda_1 &=\mathbf{G}([0,1]^\top+\mathbb{Z}^2), \\
\Lambda_2 &=\mathbf{G}([1,0]^\top+\mathbb{Z}^2), \\
\Lambda_3 &=\mathbf{G}([1,1]^\top+\mathbb{Z}^2). \\
\end{align*}
As shown in Fig. \ref{figCase}, the four cosets are marked by solid dots, solid stars,  solid triangles and solid squares, respectively. 
In addition, the packing radius of $A_2$ is
\[r_{\rm{pack}(\Lambda_f)}=1/2,\]
where the red circles  denote the largest packing spheres inside $A_2$, and the 
 cover vector $\mathbf{s}$ is marked with a cross.
 
 Depending on the given $\mathbf{m}_i$, QIM encodes $\mathbf{s}$ to a coset $\Lambda_i$. In contrast with QIM, the encoded points of MD-QIM are much closer to $\mathbf{s}$, as marked by dash circles in Fig. \ref{figCase}. For instance, to hide $\mathbf{m}_0$, $\mathbf{s}$ stays still and the process denotes a Type I encoding; to hide $\mathbf{m}_2$, $\mathbf{s}$ moves to the empty triangle inside the dash circle, and the process denotes a Type II encoding.
 
\subsection{Theoretical Analysis of Distortion}
To evaluate the distortion caused by data embedding, we define the mean square error (MSE) metric as
\begin{align}
    MSE \triangleq \frac{1}{N M} \sum_{k=1}^{M}\|\mathbf{s}_k-\mathbf{s}_{w,k}\|^2,
\end{align}
where the subscript $k$ indicates a specific host/embedded vector, and $M$ denotes the total number of host vectors. 
 
 
As each coset representative $\mathbf{d}_i$ of the coset $\Lambda_i$ ($0 \leq i \textless |\rm{det} \mathbf{J}|$) plays the role of $N$-dimensional dither vectors, the QIM method described in this paper is a type of dithered QIM. Based on the
 high-resolution
quantization assumption  \cite{zamir2014}, the quantization noise can be modeled as random and  independent of $\mathbf{s}$, and uniformly distributed over the Voronoi region of the coarse lattice. 
Then embedding distortion per dimension of the original QIM method can be measured as \cite{MoulinK05}:
\begin{align}
    MSE_{QIM}
    &= \frac{1}{N} \frac{1}{\rm{Vol}(\mathcal{V}_{\Lambda_c})}\int_{\mathcal{V}_{\Lambda_c}} \|\mathbf{x}\|^2 d\mathbf{x} \nonumber\\
    &= G(\Lambda_c){\rm{Vol}}(\mathcal{V}_{\Lambda_c})^{\frac{2}{N}},
\end{align}
in which $G(\Lambda_c)$ refers to the normalized second moment of $\Lambda_c$ and is invariant to scaling.
 Ref. \cite{conway2013sphere} has listed   the normalized second-order moments of popular low-dimensional quantizers, e.g., we have $G(\mathbb{Z})=0.083333$, $G(A_2)=0.080188$, $G(D_4)=0.076603$, $G(E_8)=0.071682$.
 
%

 {\begin{thm}\label{THM1}
		Given $\Lambda_f$ and $\Lambda_c$  {obtained via $\mathbf{G}_f$ and $\mathbf{G}_f\mathbf{J}$}, based on the
		high-resolution
		quantization assumption  \cite{zamir2014} the 
		actual MSE of MD-QIM satisfies
		\begin{align}
		&MSE_{MD{\text -}QIM} \nonumber\\
		&\geq MSE_{QIM}- \frac{1}{N}\frac{|\rm{det} \mathbf{J}|-1}{|\rm{det} \mathbf{J}|}2(r_{\rm{pack}(\Lambda_f)} - \epsilon)\times \nonumber\\
		&\quad \sqrt{NG(\Lambda_c){\rm{Vol}}(\mathcal{V}_{\Lambda_c})^\frac{2}{N}- NG(\Lambda_f){\rm{Vol}}(\mathcal{V}_{\Lambda_f})^\frac{2}{N}} \nonumber \\
		&\quad + \frac{1}{N}\frac{|\rm{det} \mathbf{J}|-1}{|\rm{det} \mathbf{J}|}(r_{\rm{pack}(\Lambda_f)} - \epsilon)^2 \nonumber \\
		&\quad - \frac{1}{|\rm{det} \mathbf{J}|}G(\Lambda_f){\rm{Vol}}(\mathcal{V}_{\Lambda_f})^{\frac{2}{N}}. \label{mselowerbound}
		\end{align}
\end{thm}}
\begin{proof}
	Define the distortion-saving metric as \begin{equation}
	D= MSE_{QIM} - MSE_{MD{\text -}QIM}.
	\end{equation}
We need to prove that
	\begin{align}
	D &\leq \frac{1}{N}\frac{|\rm{det} \mathbf{J}|-1}{|\rm{det} \mathbf{J}|}2(r_{\rm{pack}(\Lambda_f)} - \epsilon)\times \nonumber\\
	&\quad \sqrt{NG(\Lambda_c){\rm{Vol}}(\mathcal{V}_{\Lambda_c})^\frac{2}{N}-NG(\Lambda_f){\rm{Vol}}(\mathcal{V}_{\Lambda_f})^\frac{2}{N}} \nonumber \\
	&\quad - \frac{1}{N}\frac{|\rm{det} \mathbf{J}|-1}{|\rm{det} \mathbf{J}|}(r_{\rm{pack}(\Lambda_f)} - \epsilon)^2 \nonumber \\
	&\quad + \frac{1}{|\rm{det} \mathbf{J}|}G(\Lambda_f){\rm{Vol}}(\mathcal{V}_{\Lambda_f})^{\frac{2}{N}}. \label{upperbound}
	\end{align}
	 {
		Due to the  high-resolution
		quantization assumption  \cite{zamir2014}, $\mathbf{p}_i$ can be modeled as a generalized dither admitting a uniform distribution. Thus
		\begin{equation}
		\mathrm{Pr}(\mathbf{p}_i \in \mathcal{V}_{\Lambda_f})=\frac{1}{|\rm{det} \mathbf{J}|},\,  	\mathrm{Pr}(\mathbf{p}_i \in \mathcal{V}_{\Lambda_c} \backslash \mathcal{V}_{\Lambda_f})=\frac{|\rm{det} \mathbf{J}|-1}{|\rm{det} \mathbf{J}|}.
		\end{equation}
		Let $D_1$, $D_2$ be the distortion saving caused by the cases of $\mathbf{p}_i \in \mathcal{V}_{\Lambda_f}$ and $\mathbf{p}_i \in \mathcal{V}_{\Lambda_c} \backslash \mathcal{V}_{\Lambda_f}$.  The overall distortion saving can be formulated as
		\begin{equation}\label{prof_total}
		D= \frac{1}{|\rm{det} \mathbf{J}|}D_1 + \frac{|\rm{det} \mathbf{J}|-1}{|\rm{det} \mathbf{J}|}D_2.
		\end{equation}
		For the scenario of $\mathbf{p}_i \in \mathcal{V}_{\Lambda_f}$,  based on Eq. (\ref{staystill}) we have
		\begin{equation} 
		D_1 \leq \mathbb{E}[\|\mathcal{P}\|^2]/N = {\rm{Vol}}(\mathcal{V}_{\Lambda_f})^{\frac{2}{N}}G(\Lambda_f),
		\end{equation}
		where $\mathcal{P}$ denotes a random vector uniformly distributed over $\mathcal{V}_{\Lambda_f}$. This upper bound is tight if the fine lattice is good for packing.}
		
		 {For the scenario of $\mathbf{p}_i \in \mathcal{V}_{\Lambda_c} \backslash \mathcal{V}_{\Lambda_f}$, 
			{note} that  our difference vector is
			\begin{equation} \label{eq_inTypeii}
			\mathbf{p}_i'  \triangleq \mathbf{p}_i - 
			\frac{\mathbf{p}_i}{\| \mathbf{p}_i \|}\times(r_{\rm{pack}(\Lambda_f)} - \epsilon).
			\end{equation}
			Obviously  $\mathbf{p}_i'$ is in the same direction as $\mathbf{p}_i$ while $\| \mathbf{p}_i' \| < \| \mathbf{p}_i \|$.  Since
			\begin{equation}
			D_2=( \mathbb{E}[\|\mathbf{p}_i\|^2] - \mathbb{E}[\|\mathbf{p}_i'\|^2])/N,
			\end{equation}
			by substituting $\mathbf{p}_i'$ with Eq. (\ref{eq_inTypeii}), we have
			\begin{equation}
			D_2 = \frac{1}{N}( 2 (r_{\rm{pack}(\Lambda_f)} - \epsilon) \mathbb{E} [\| \mathbf{p}_i \|] -  
			(r_{\rm{pack}(\Lambda_f)} - \epsilon)^2).
			\end{equation}}
		 {To construct the upper bound for $\mathbb{E}[\|\mathbf{p}_i\| ]$,
			it follows from the Cauchy–Schwarz inequality 
			 that  
			\begin{align}
			&\mathbb{E}[\|\mathbf{p}_i\| ] \nonumber\\
			 &= \frac{1}{{\rm{Vol}}(\mathcal{V}_{\Lambda_c} \backslash \mathcal{V}_{\Lambda_f})} \int_{\mathbf{u}\in \mathcal{V}_{\Lambda_c} \backslash \mathcal{V}_{\Lambda_f} }  ||\mathbf{u}|| \mathrm{d}\mathbf{u} \nonumber \\
			& \leq \frac{1}{{\rm{Vol}}(\mathcal{V}_{\Lambda_c} \backslash \mathcal{V}_{\Lambda_f})} \sqrt{ \int_{\mathbf{u}\in \mathcal{V}_{\Lambda_c} \backslash \mathcal{V}_{\Lambda_f} }  ||\mathbf{u}||^2 \mathrm{d}\mathbf{u}} \sqrt{{\rm{Vol}}(\mathcal{V}_{\Lambda_c} \backslash \mathcal{V}_{\Lambda_f})} \nonumber \\
			&=  \sqrt{NG(\Lambda_c){\rm{Vol}}(\mathcal{V}_{\Lambda_c})^{\frac{2}{N}} - 
				NG(\Lambda_f){\rm{Vol}}(\mathcal{V}_{\Lambda_f})^{\frac{2}{N}}}.
			\end{align}
			By substituting the above inequalities to Eq. (\ref{prof_total}), the upper bound of $D$ is derived and the theorem holds.}
\end{proof}

The power of Theorem 1 is that, as our simulation results will show, the lower bound often accurately predicts the actual MSE of MD-QIM.
 
\subsection{ {Remarks}}
 {Regarding computational complexity, the cost of MD-QIM is essentially the same as that of QIM, as they both employ only one nearest neighbor search over the coarse lattice $\Lambda_c$. The additional cost of MD-QIM is minor, where $||\mathbf{p}_i||$ consumes $2N$ floating point operations (FLOPS), and the division and subtraction in Eq. (\ref{cost_mdqim_2}) consume $2N$ FLOPS.}

This work omits the usage of additional random dithers for two reasons. Firstly, \textit{Wang et al.}  \cite{yuangen21} have shown that using secret dithers as keys can be easily defeated based on known message attacks. To impose strong security on QIM/MD-QIM, symmetric cryptography can be applied to the messages before their embedding. The objective hereby is to meet the requirement of ``getting unnoticed'' for steganography. Secondly, dithering would increase   distortion when the code rate is low \cite{zamir2014}.

 {
There are a large variety of state-of-the-art methods which allow   particular constraints (transparency, robustness, data payload, etc) to be individually reached for different applications. However, no single method can jointly reach all the requirements.  Compared to QIM, the robustness of MD-QIM is sacrificed to achieve the minimum amount of distortion which lead to high transparency property. The payload of MD-QIM is the same as that of QIM, both equal to $|\rm{det} \mathbf{J}|$.} 

\section{Simulations}
We carry out numerical simulations in this section to verify the effectiveness of MD-QIM. 
We have chosen the 
MIT-BIH arrhythmia database \cite{moody2001impact} as our  cover ECG sources, which contains   $48$ ECG records labeled as No. 1-48.  
Random messages admitting uniform distributions are taken as embedded sources.  { For illustrative purposes, we employ the $\mathbb{Z}$, $A_2$,  {$D_4$ and $E_8$} lattices as $\Lambda_f$,   respectively.  Their corresponding coarse lattices $\Lambda_c$ are set as  $\alpha \Lambda_f$, 
where   $\alpha$ denotes a parameter that  controls the embedding rate of messages ($R= \log \alpha$).}   {
Benchmark algorithms are taken from \cite{MoulinK05}  and \cite{ibaida2011low,rmit1}. The method in \cite{MoulinK05} represents a simple and efficient implementation of QIM, while  the method in \cite{ibaida2011low,rmit1} denotes an efficient alternative method to achieve information hiding with minimum distortion.} 
 


In addition to the aforementioned MSE metric, we introduce two more popular metrics used in data hiding literature  to measure the imperceptibility of MD-QIM. The peak signal to noise ratio (PSNR) is defined as
\begin{align}
&PSNR = \nonumber \\
& 20\times\log_{10}
\frac{\max\limits_{k\in\{1,...,M\}}(\max(\mathbf{s}_k))-\min\limits_{k\in\{1,...,M\}}(\min(\mathbf{s}_k))}{\sqrt{MSE}}.
\end{align}
The percentage residual difference (PRD) is given by
\begin{align}
    PRD = \sqrt{\frac{\sum_{i=1}^{M}\|\mathbf{s}_k-\mathbf{s}_{w,k}\|^2}{\sum_{i=1}^{M}\|\mathbf{s}_k\|^2}} \times 100\%.
\end{align}

\begin{table*}[t!]
\centering
\renewcommand\arraystretch{1.5}
\caption{ {Comparisons with benchmarks.}}  
\label{experimentResult}  
\begin{tabular}{|c|c|c|c|c|c|c|c|c|c|c|}
\hline
\multirow{2}{*}{\textbf{Rate} $R$} & \multirow{2}{*}{\textbf{Metrics}} & \multicolumn{4}{c|}{\textbf{Original QIM} \cite{MoulinK05}} & \multicolumn{4}{c|}{\textbf{Proposed MD-QIM}} & \multicolumn{1}{c|}{\multirow{2}{*}{\begin{tabular}[c]{@{}c@{}}\textbf{Joint Wavelet and}  \\ \textbf{Scalar QIM} \cite{ibaida2011low,rmit1}\end{tabular}}} \\ \cline{3-10}
& & $\mathbb{Z}$ & $A_2$ & $D_4$ & $E_8$ & $\mathbb{Z}$ & $A_2$ & $D_4$ & $E_8$ & \multicolumn{1}{c|}{}        
\\ \hline
\multicolumn{1}{|c|}{\multirow{4}{*}{1}} & MSE  
& 0.1792 & 0.1718 & 0.1659 & 0.1553 
& 0.0238 & 0.0304 & 0.0411 & 0.0467 
& 0.1103               
\\ \cline{2-11} 

& PSNR  (dB)
& 63.58 & 63.76 & 63.91 &  64.20 
& 72.35 & 71.28 & 69.97 & 69.42 
& 65.68 
\\ \cline{2-11} 

& PRD 
&0.37\% & 0.36\%  & 0.35\%  & 0.34\% 
&0.13\%  & 0.15\%  & 0.17\%  & 0.19\% 
&0.29\% 
\\ \hline

\multicolumn{1}{|c|}{\multirow{4}{*}{2}} & MSE
&0.7172&0.6842&0.6591&0.6173
&0.3088&0.3328&0.3650&0.3682
&0.5493 
\\ \cline{2-11} 

 & PSNR (dB)
& 57.55& 57.76& 57.92& 58.20
& 61.21 & 60.89 & 60.49 & 60.45
& 58.71
 \\ \cline{2-11} 

& PRD      
 & 0.73\%& 0.71\%& 0.70\%& 0.68\%
 & 0.48\%& 0.50\%& 0.52\%& 0.52\% 
 & 0.64\%
\\ \hline
\end{tabular}
\end{table*}

 {In
	Table \ref{experimentResult}, we list the {average values of} MSE, PSNR and PRD  {of 48 samples} on QIM, MD-QIM  {and wavelet-based ECG steganography, respectively}. Some observations can be made from the table. i) MD-QIM outperforms QIM  {and the Wavelet-based ECG steganography} in all the metrics (MSE, PSNR and PRD), regardless of the chosen type of lattices. For instance the MSE of MD-QIM in  {lattice $A_2$} ($R=1$) is  {$0.0304$}, which costs only around $20\%$ of the $0.1718$ MSE in QIM. ii) As the code rate $R$ increases, the MSE/PRD also rises, due to the fact that we have fixed the fine lattices to feature unit volumes while the volume of the coarse lattices grow as $R$ increases. iii) Choosing large-dimensional good lattices is much desirable. Steady improvements can be observed by increasing the dimensions of the lattices from $\mathbb{Z}$ to $A_2$, $D_4$ and $E_8$.} 

\begin{figure*}[ht]
\centering
\subfigure[Results based on the $A_2$ lattice.]{\includegraphics[width=5.9cm]{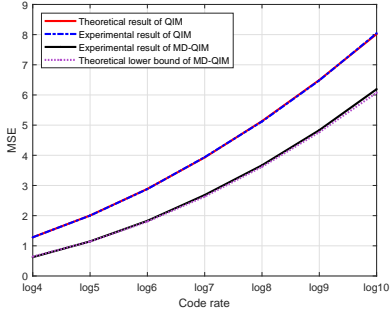}} 
\subfigure[Results based on the $D_4$ lattice.]{\includegraphics[width=5.9cm]{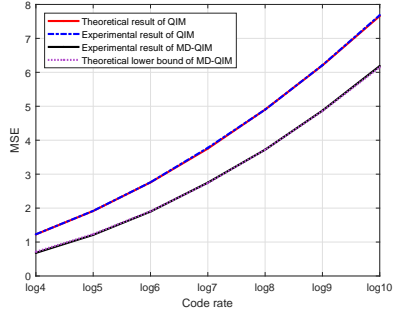}}
\subfigure[Results based on the $E_8$ lattice.]{\includegraphics[width=5.9cm]{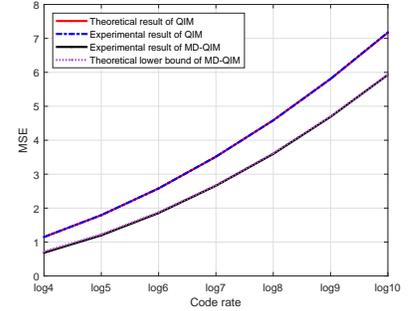}}
\caption{ {The theoretical and simulated MSE values of MD-QIM and QIM based on different lattices.}} 
\label{figTheory}
\end{figure*}

  {To justify the accuracy of Theorem \ref{THM1}, based on the $A_2$, {$D_4$ and $E_8$} lattices, 
 	we plot the theoretical and simulated MSE values of both MD-QIM and QIM  {in Fig. \ref{figTheory}}. All the subfigures justify the tightness of  our lower bound in Eq. (\ref{mselowerbound}), whose values are very close to the actual simulated ones.  
 	Therefore, the amount of reduced MSE caused by MD-QIM is indeed well characterized by Theorem 1.}

\section{Conclusions}
In this letter a lattice-based minimum distortion data hiding method has been proposed, which is particularly suitable for hiding data in some applications where the robustness to the AWGN attack is not the primary concern. Rigorous analysis of the distortion saving has been presented, and the analysis is universal in terms of the size and type of lattice bases.  Simulations show that the proposed method features a much smaller amount of distortion compared with the conventional QIM  {and Wavelet-based ECG steganography}.

\bibliographystyle{IEEEtranMine}
\bibliography{lib}

\begin{thebibliography}{10}
\providecommand{\url}[1]{#1}
\csname url@samestyle\endcsname
\providecommand{\newblock}{\relax}
\providecommand{\bibinfo}[2]{#2}
\providecommand{\BIBentrySTDinterwordspacing}{\spaceskip=0pt\relax}
\providecommand{\BIBentryALTinterwordstretchfactor}{4}
\providecommand{\BIBentryALTinterwordspacing}{\spaceskip=\fontdimen2\font plus
\BIBentryALTinterwordstretchfactor\fontdimen3\font minus
  \fontdimen4\font\relax}
\providecommand{\BIBforeignlanguage}[2]{{%
\expandafter\ifx\csname l@#1\endcsname\relax
\typeout{** WARNING: IEEEtran.bst: No hyphenation pattern has been}%
\typeout{** loaded for the language `#1'. Using the pattern for}%
\typeout{** the default language instead.}%
\else
\language=\csname l@#1\endcsname
\fi
#2}}
\providecommand{\BIBdecl}{\relax}
\BIBdecl

\bibitem{cox99}
I.~J. Cox, M.~L. Miller, and A.~L. McKellips, ``Watermarking as communications
  with side information,'' \emph{Proceedings of the IEEE}, vol.~87, no.~7, pp.
  1127--1141, 1999.

\bibitem{zamir2014}
R.~Zamir, \emph{Lattice Coding for Signals and Networks}.\hskip 1em plus 0.5em
  minus 0.4em\relax Cambridge: Cambridge University Press, 2014.

\bibitem{lingtit14}
\BIBentryALTinterwordspacing
C.~Ling, L.~Luzzi, J.~Belfiore, and D.~Stehl{\'{e}}, ``Semantically secure
  lattice codes for the gaussian wiretap channel,'' \emph{{IEEE} Trans. Inf.
  Theory}, vol.~60, no.~10, pp. 6399--6416, 2014.
\BIBentrySTDinterwordspacing

\bibitem{feng16a}
\BIBentryALTinterwordspacing
B.~Feng, W.~Lu, W.~Sun, J.~Huang, and Y.~Shi, ``Robust image watermarking based
  on tucker decomposition and adaptive-lattice quantization index modulation,''
  \emph{Signal Process. Image Commun.}, vol.~41, pp. 1--14, 2016.
\BIBentrySTDinterwordspacing

\bibitem{feng16b}
\BIBentryALTinterwordspacing
B.~Feng, J.~Weng, W.~Lu, and B.~Pei, ``Multiple watermarking using multilevel
  quantization index modulation,'' in \emph{Digital Forensics and Watermarking
  - 15th International Workshop, {IWDW} 2016, Beijing, China}, pp. 312--326,
  2016.
\BIBentrySTDinterwordspacing

\bibitem{changshengtip19}
\BIBentryALTinterwordspacing
C.~Chen, W.~Huang, L.~Zhang, and W.~H. Mow, ``Robust and unobtrusive
  display-to-camera communications via blue channel embedding,'' \emph{{IEEE}
  Trans. Image Process.}, vol.~28, no.~1, pp. 156--169, 2019.
\BIBentrySTDinterwordspacing

\bibitem{ChenW01}
\BIBentryALTinterwordspacing
B.~Chen and G.~W. Wornell, ``Quantization index modulation: {A} class of
  provably good methods for digital watermarking and information embedding,''
  \emph{{IEEE} Trans. Inf. Theory}, vol.~47, no.~4, pp. 1423--1443, 2001.
\BIBentrySTDinterwordspacing

\bibitem{MoulinK05}
\BIBentryALTinterwordspacing
P.~Moulin and R.~Koetter, ``Data-hiding codes,'' \emph{Proc. {IEEE}}, vol.~93,
  no.~12, pp. 2083--2126, 2005.
\BIBentrySTDinterwordspacing

\bibitem{e803}
Q.~Zhang and N.~Boston, ``Quantization index modulation using the {E8}
  lattice,'' in \emph{Proceedings of the Annual Allerton Conference on
  Communication Control and Computing}, no.~1, pp. 488--489, 2003.

\bibitem{conway2013sphere}
J.~H. Conway and N.~J.~A. Sloane, \emph{Sphere packings, lattices and
  groups}.\hskip 1em plus 0.5em minus 0.4em\relax Springer Science \& Business
  Media, 2013, vol. 290.

\bibitem{OuriqueLJP05}
\BIBentryALTinterwordspacing
F.~Ourique, V.~Licks, R.~Jordan, and F.~P{\'{e}}rez{-}Gonz{\'{a}}lez, ``Angle
  {QIM:} a novel watermark embedding scheme robust against amplitude scaling
  distortions,'' in \emph{2005 {IEEE} International Conference on Acoustics,
  Speech, and Signal Processing, {ICASSP} '05, Philadelphia, Pennsylvania,
  USA}, pp. 797--800, 2005.
\BIBentrySTDinterwordspacing

\bibitem{KalantariA10}
\BIBentryALTinterwordspacing
N.~K. Kalantari and S.~M. Ahadi, ``A logarithmic quantization index modulation
  for perceptually better data hiding,'' \emph{{IEEE} Trans. Image Process.},
  vol.~19, no.~6, pp. 1504--1517, 2010.
\BIBentrySTDinterwordspacing

\bibitem{ibaida2011low}
A.~Ibaida, I.~Khalil, and R.~Van-Schyndel, ``A low complexity high capacity ecg
  signal watermark for wearable sensor-net health monitoring system,'' in
  \emph{2011 Computing in Cardiology}, pp. 393--396, 2011.

\bibitem{kaur2010digital}
S.~Kaur, R.~Singhal, O.~Farooq, and B.~S. Ahuja, ``Digital watermarking of
  {ECG} data for secure wireless commuication,'' in \emph{International
  Conference on Recent Trends in Information, Telecommunication and Computing},
  pp. 140--144, 2010.

\bibitem{rmit1}
\BIBentryALTinterwordspacing
A.~Ibaida and I.~Khalil, ``{Wavelet-Based {ECG} Steganography for Protecting
  Patient Confidential Information in Point-of-Care Systems},'' \emph{{IEEE}
  Trans. Biomed. Eng.}, vol.~60, no.~12, pp. 3322--3330, 2013.
\BIBentrySTDinterwordspacing

\bibitem{lyutit19}
\BIBentryALTinterwordspacing
S.~Lyu, A.~Campello, and C.~Ling, ``Ring compute-and-forward over block-fading
  channels,'' \emph{{IEEE} Trans. Inf. Theory}, vol.~65, no.~11, pp.
  6931--6949, 2019.
\BIBentrySTDinterwordspacing

\bibitem{yuangen21}
\BIBentryALTinterwordspacing
Y.~Wang, G.~Zhu, J.~Li, M.~Conti, and J.~Huang, ``Defeating lattice-based data
  hiding code via decoding security hole,'' \emph{{IEEE} Trans. Circuits Syst.
  Video Technol.}, vol.~31, no.~1, pp. 76--87, 2021.
\BIBentrySTDinterwordspacing

\bibitem{moody2001impact}
G.~B. Moody and R.~G. Mark, ``The impact of the {MIT-BIH} arrhythmia
  database,'' \emph{IEEE Engineering in Medicine and Biology Magazine},
  vol.~20, no.~3, pp. 45--50, 2001.

\end{thebibliography}

\end{document}